# Optical feedback induced irregular and Chaotic dynamics in terahertz quantum cascade laser combs

Xiaoqiong Qi,[1] Carlo Silvestri,[1, 2] Thomas Taimre,[3] and Aleksandar D. Rakić[1]

[1])*School of Electrical Engineering and Computer Science, The University of Queensland, Brisbane, QLD 4072 Australia*

[2])*Institute of Photonics and Optical Science (IPOS), School of Physics, The University of Sydney, NSW 2006, Australia*

[3])*School of Mathematics and Physics, The University of Queensland, Brisbane, QLD 4072 Australia*

(*Electronic mail: * a.rakic@uq.edu.au; x.qi1@uq.edu.au.)

In this work, we systematically investigate the optical feedback dynamics of terahertz (THz) frequency combs generated from quantum cascade lasers (QCLs) using the effective semiconductor Maxwell–Bloch equations (ESMBEs). Starting from a free-running comb state at low bias ($1.3 I_{th}$), we identify clear bifurcation routes as the external feedback is increased. In the weak-feedback regime ($C < 1$), the frequency comb operation remains stable; as feedback increases to the moderate feedback regime ($1 < C < 2.4$), the system transitions through frequency splitting into period-one (P1) oscillations around each comb, followed by higher-order bifurcations (P2, P4, P8). Within the range $2.4 < C < 13.86$, only P1 states are observed, with the split frequency increasing continuously with feedback strength. In the strong feedback regime ($C > 13.86$), complex feedback dynamics re-emerged but with significantly reduced complexity compared with the weak feedback region. More interestingly, we show that optical feedback can induce chaotic THz emission at higher bias current conditions ($> 1.6 I_{th}$), when the free-running laser output ($C = 0$) is originally a stable frequency comb. The resulting chaotic dynamics, characterized by broadband spectra and large positive Lyapunov exponents, are shown to depend strongly on the linewidth enhancement factor. Although THz QCLs are widely considered as ultrastable under external perturbations, this work provides the first detailed report of chaotic dynamics and routes to chaos in THz QCLs under optical feedback. These findings not only deepen the understanding of nonlinear dynamics of THz QCLs but also open new possibilities for exploiting applications of chaotic light in THz sensing, imaging and secure communications.

## I. INTRODUCTION

Semiconductor lasers exhibit a wide range of nonlinear dynamical behaviors, with chaotic dynamics being among the most intriguing phenomena[1,2]. Such instabilities typically arise from the nonlinear interplay between wave propagation within the laser cavity and radiative recombination, which produces a macroscopic polarization encapsulated in both the polarization field and the carrier inversion[2,3]. Chaos were generally induced through configurations involving optical feedback[4–9], optical injection[10–13], pump-current modulation[14,15], optoelectronic feedback[16,17], or even mode competition without external perturbation[18,19].

Chaotic behavior in semiconductor lasers has been extensively studied not only for its fundamental nonlinear dynamics but also for its practical applications in random number generation, secure communication, chaos computing, remote sensing, and imaging cryptography[20–28]. However, most research on chaotic semiconductor lasers has concentrated on near-infrared and mid-infrared devices—such as laser diodes (LDs)[5,7,8,19,24,26,28], and mid-infrared quantum cascade lasers (QCLs)[9,29]. In contrast, chaotic emission from far-infrared or terahertz (THz) semiconductor lasers, such as THz QCLs has been rarely explored. Chaos was recently reported in free running multimode THz QCLs[30] but it has never been systematically investigated in presence of optical feedback. Studying the irregular and chaotic dynamics in THz QCLs under optical feedback opens new opportunities for THz sensing, imaging and spectroscopy application (such as THz chaos LiDAR), as well as secure communications over wide-bandwidth THz channels. Moreover, such investigation are essential for identifying the stable pulse generation regimes required for THz frequency combs applications.

THz QCLs have been believed to be highly stable under delayed optical feedback due to the absence of relaxation oscillations and lower linewidth enhancement factor than those of LDs and MIR-QCLs[31,32]. However, recent studies challenge this assumption: self-pulsations have been experimentally demonstrated and theoretically explored in THz QCLs under optical feedback[33], with oscillation frequency increasing with increasing feedback strength. The self-pulsation dynamics existing in intrinsic THz QCLs can interact with the self-mixing induced by driving current sweeping and produce more complex dynamic waveforms[34]. The self-pulsation phenomenon has also been applied for THz imaging applications[35] and is highly suitable for high temperature THz QCLs with very limited driving current operation range[36]. Tilted feedback has also been reported in THz QCLs to generate periodic, quasi-periodic oscillations, and square-wave patterns[37]. Optical feedback can also trigger fundamental and harmonic frequency combs in QCLs[38]. Although various feedback dynamics has been reported for THz QCLs, chaotic light and the related routes to chaos have not been reported in THz QCLs under optical feedback.

The chaotic dynamics in LDs under optical feedback typically results from the competition between the laser's intrinsic relaxation oscillation frequency ($f_{RO}$) and the resonant frequency of the external cavity ($f_{EC}$)[5,6,8,39]. Further study shows that LDs exhibits chaotic behavior only in the long-cavity regime when $f_{RO} \gg f_{EC}$. In the short-cavity regime when $f_{RO} < f_{EC}$, periodic pulsing and regular pulse packets replace chaos[7]. Depending on the delay ratio of the external cavity delay time and the inverse of the $f_{RO}$ of the laser, different routes to chaos have been reported, which includes the quasi-periodic route, subharmonic oscillation route, and periodic doubling route[6]. Amongst these, the quasi-periodic route is a universal crisis route to chaos in LDs subject to external optical feedback that involves these two fundamental frequencies $f_{RO}$ and $f_{ext}$[39]. However, there are no relaxation oscillations in THz QCLs due to their ultrafast intersubband scattering and very short upper-state lifetime that is comparable to the photon lifetime on the picosecond scale. This prevent the build-up and depletion of a carrier reservoir needed to sustain oscillatory carrier–photon coupling. Therefore, the route to chaos in this type of laser are still unknown.

In this work, we have systematically explored optical feedback effects on a THz QCL with a frequency comb output. Interestingly, a period doubling route into chaos has been found through RF beat note spectrum with ultra-short step size of the feedback reflectivity (on the order of 0.001%). The modeling of the optical feedback strength is based on the effective Maxwell Bloch equations. The finding of chaotic light output from a THz QCL is in contrast to the generally accepted claim that THz QCLs are ultrastable against feedback[31,32]. It was found a periodic doubling route starting from a excited period-one oscillation which is 10 times smaller than the external cavity mode $f_{ext}$ but gradually increasing to $f_{ext}$ with growing external reflectivity. This frequency modulation properties can be explained by the excess phase equations[33]. Optical feedback triggers chaos at high bias current conditions ( $> 1.6$ $I_{th}$) when the laser output is originally frequency comb at free running. The complexity of chaos and bandwidth can be improved by higher linewidth enhancement factor, evidenced by the increasing largest Lyapunov exponent and the power spectrum, respectively. The proposed chaotic dynamics is crucial for advancing applications in sensing, imaging, spectroscopy, and secure communication technologies in the THz frequency band.

## II. THEORETICAL MODEL OF THZ FREQUENCY COMBS WITH OPTICAL FEEDBACK

To investigate the frequency comb dynamics of THz-QCLs under optical feedback, we employ the effective semiconductor Maxwell–Bloch equations (ESMBEs), originally introduced in[40] to describe the multimode behavior of MIR QCLs in a Fabry–Perot (FP) configuration. The ESMBE framework has proven to be effective in capturing the key features of self-generated OFCs in both free running and fundamental frequency combs regions. It is based on a phenomenological formulation of the semiconductor material's optical susceptibility and incorporates key physical effects, including the linewidth enhancement factor, the carrier-density dependence of gain and refractive index, and spatial hole burning, which models the evolution of a carrier grating within the FP cavity.

The complete set of ESMBEs used in this study is as follows:

$$\frac{\partial E^+}{\partial z} + \frac{1}{v}\frac{\partial E^+}{\partial t} = -\frac{\alpha_L}{2}E^+ + gP_0^+ \tag{1}$$

$$-\frac{\partial E^-}{\partial z} + \frac{1}{v}\frac{\partial E^-}{\partial t} = -\frac{\alpha_L}{2}E^- + gP_0^- \tag{2}$$

$$\frac{\partial P_0^+}{\partial t} = \pi\delta_{hom}(1+i\alpha)\left[-P_0^+ + if_0\varepsilon_0\varepsilon_b(1+i\alpha)(N_0E^+ + N_1^+E^-)\right] \tag{3}$$

$$\frac{\partial P_0^-}{\partial t} = \pi\delta_{hom}(1+i\alpha)\left[-P_0^- + if_0\varepsilon_0\varepsilon_b(1+i\alpha)(N_0E^- + N_1^-E^+)\right] \tag{4}$$

$$\frac{\partial N_0}{\partial t} = \frac{I}{eV} - \frac{N_0}{\tau_e} + \frac{i}{4\hbar}\left[E^{+*}P_0^+ + E^{-*}P_0^- - E^+P_0^{+*} - E^-P_0^{-*}\right] \tag{5}$$

$$\frac{\partial N_1^+}{\partial t} = -\frac{N_1^+}{\tau_e} + \frac{i}{4\hbar}\left[E^{-*}P_0^+ - E^+P_0^{-*}\right] \tag{6}$$

where $E^+(z,t)$ and $E^-(z,t)$ are the forward and backward envelopes for the electric fields propagating inside the laser cavity, $P_0^+$, $P_0^-$ are the forward and backward polarization envelopes, $N_0$ is the density of carriers at zero-order. $N_1^+$ and $N_1^-$ are the first-order carrier density terms, which reproduce the carrier grating due to spatial hole burning occurring in the FP resonator. The other parameters and their values are defined in Table I. The optical feedback effect on QCL emission can be described by modifying the boundary conditions for the FP cavity [Eq. (7) and (8)][38], where $L$ and $L_{ext}$ denote the laser cavity length and the external cavity length, respectively. The target reflectivity $R_{ext}$ enters through the total coupling coefficient $\varepsilon = R_{ext}\varepsilon_L\varepsilon_S$, where $\varepsilon_L$ and $\varepsilon_S$ are coupling factors that describe the losses in the external cavity and those associated with the reinjection into the laser cavity. In this work, we assume $\varepsilon_L = \varepsilon_S = 1$ and we vary $R_{ext}$ from 0 to 100% to represent different optical feedback levels.

$$E^-(L,t) = \sqrt{R}E^+(L,t) + \varepsilon t_L^2 E^+(L,t-\tau_{ext})exp(-i\omega_0\tau_{ext}), \tag{7}$$

$$E^+(0,t) = \sqrt{R}E^-(0,t). \tag{8}$$

## III. THZ QCL COMBS UNDER OPTICAL FEEDBACK

To illustrate how the laser dynamics of THz QCL frequency combs evolve under optical feedback, we divided our simulations into distinct regions based on the target reflectivity range, $R_{ext}$. The most interesting dynamics occur in the range $R_{ext} = 0 - 2\%$, which we examine first, follwed by an analysis of the remaining ranges.

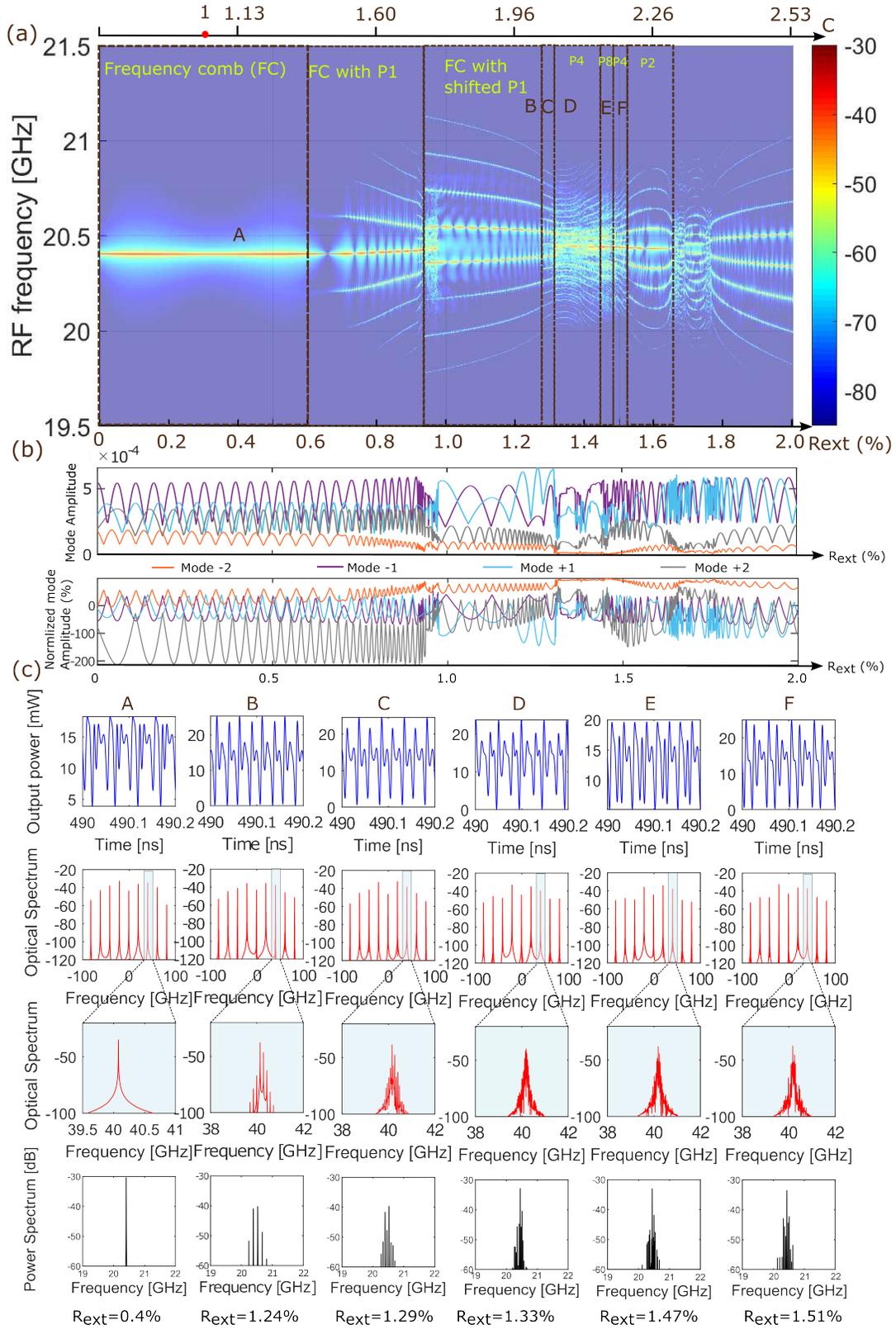

FIG. 1. Optical feedback dynamics transition as a function of the external reflectivity $R_{ext}$ varying from 0–2.0% : (a) RF beat note map; (b) Mode amplitude variations for the dominant mode: Mode -2, Mode -1, Mode +1, and Mode +2; (c) Time traces, optical output spectra, and power spectrum at representative feedback levels from A to E as labeled in (a).

We present in Fig. 1(a) the RF beat-note spectrum map around the laser's FSR (20.4 GHz), together with the dominant optical longitudinal mode amplitude variations [Fig. 1 (b)], the corresponding time traces, optical output spectra, and power spectrum at some representative target reflectivity $R_{ext}$ values [Fig. 1 (c)] when $R_{ext}$ increases from 0 to 2%. We observe a clear dynamics transitions: from pure frequency comb (FC) operation to each FC line with an excited single-frequency Period-One (P1) oscillations, then to an FC with a shifted P1 state. This is followed by a further spectral slitting at half the excited frequency, corresponding to Period-two (P2) state, and further bifurcation leading to Period-four (P4) and Period-eight (P8) states and quasi-chaotic state. The laser output returns to P4, P2 states after that. Typical laser output time traces, optical spectra, zoomed-in optical spectrum at Mode +2 (around 40 GHz), and the power spectrum at $R_{ext}$=0.4% (A), 1.24% (B), 1.29% (C), 1.33% (D), 1.47% (E), and 1.51% (F) are shown in Fig. 1(c). Point A corresponds to a region of stable FC where each longitudinal mode amplitude varies smoothly with increasing $R_{ext}$ in Fig. 1(b) and both longitudinal modes and the power spectrum exhibit single line in Fig. 1(c) owing to the narrow linewidth of each mode and uniform mode spacing of the comb. Starting right after 0.6% reflection, frequency splitting appears in each longitudinal mode, the beat note spectrum (power spectrum) then exhibits multiple frequency components, which defined here as FC with P1. This frequency split structure shifts with increasing $R_{ext}$, as seen in the region around Point B. In the region around Point C, further frequency splitting at half of the P1 frequency occurs, producing doubled frequency components in the optical power spectrum (P2); successive doublings emerge at Point D (P4) and Point E (P8). Beyond $R_{ext} > 1.52\%$, the dynamics revert to P2 and repeat, as evident in the RF spectrum map. Notably, the P1 split frequency is around 0.15 GHz, which is 10 times smaller than the resonant external cavity frequency $f_{ext}$ of 1.5 GHz with $L_{ext}$ at 0.1 m. As shown in Figs. 3, 4, and 6, this split frequency increases continuously with $R_{ext}$. This frequency modulation property has been observed in our previous work for single-mode THz QCL under optical feedback as well[33,35], which can be well explained by the excess phase equation, where the emission frequency from the laser depends on the feedback level.

Another interesting observation is that the dominant longitudinal mode amplitudes oscillates periodically with $R_{ext}$ in the range of $0 - 0.9\%$ in Fig. 1(b), which is a behavior not pronounced in single-mode THz QCLs under optical feedback. In single-mode lasers with weak feedback, the emission mode power is proportional to the feedback coupling coefficient $\kappa$ which is in turn proportional to $\sqrt{R_{ext}}$ under weak feedback. This relationship has been exploited for THz sensing and imaging[35,41]. Here for multi-mode THz QCLs, however, the dominant mode powers oscillate with $R_{ext}$ in the regime where the laser still operates as a frequency comb. This phenomenon only became evident when the $R_{ext}$ step size was reduced to 0.001% (i.e., 2000 simulation points from 0 to 2% in Fig. 1). This mode-dependent oscillation is likely due to mode competition combined with mode-specific feedback phases: some modes experience constructive feedback while others receive destructive feedback, leading to different oscillation frequencies. The oscillating phase is also slightly different among modes due to gain competition in the laser cavity. For these modes, the $R_{ext}$ oscillation period is on the order of 0.05% differs slightly from mode to mode due to their distinct frequencies. To quantify how much the emission power varies with increasing $R_{ext}$, the emission power at each dominant mode in top panel of Fig. 1 (b) was normalized to its power without optical feedback for each mode and the results are shown in the bottom panel. The mode-amplitude variations exceed 50% for all four modes. As $R_{ext}$ increases from 0.9% to 2.0%, the previously periodic oscillations become irregular due to frequency splitting and power transfer into split modes, but since in the plot of Fig. 1(b) we tracked only the amplitude of single central frequency for each comb line at –40 GHz (Mode -2), –20 GHz (Mode -1), 20 GHz (Mode +1), and 40 GHz (Mode +2).

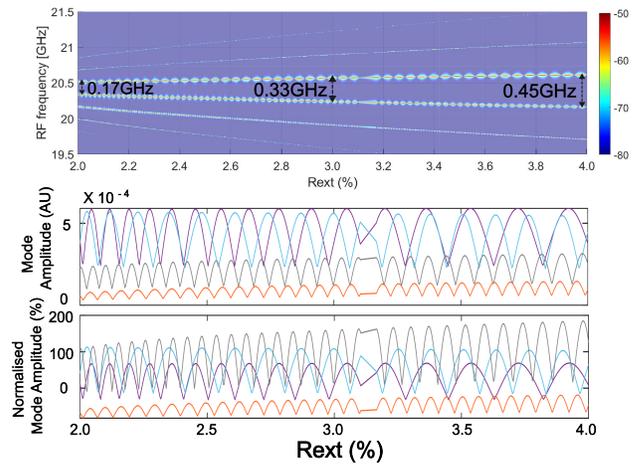

FIG. 2. RF Beat note map (top) and the mode amplitude variations (bottom) as a function of the external reflectivity $R_{ext}$, which is varied from 2.0% to 4.0%. The corresponding $C$ parameter ranges from 2.53 to 3.58.

The optical feedback dynamics for $R_{ext}$ between $2\% - 4\%$ are shown in Fig. 2. In this range, only FC with P1 oscillations are observed, with the split frequency increasing from 0.17 GHz to 0.45 GHz as the feedback strength grows. A similar trend appears in Fig. 3 for $R_{ext}$ ranging from 4% to 7%, where the split frequency continues to rise to 0.73 GHz. This behavior persists across a wide range of feedback strengths ($2\% - 60\%$), as illustrated in Fig. 4, with the split frequency monotonically increasing from 0.17 GHz all the way up to 1.29 GHz at $R_{ext} = 60\%$.

To illustrate the typical laser output across this wide range, Fig.5 compares the behavior of the free running [Fig. 5 left] with that under $R_{ext}$ is 30% [Fig. 5 right]. The most significant difference appears in the optical spectra, where feedback introduces subcombs at each comb line with a frequency spacing of 1.18 GHz. Correspondingly, in the time domain, the pulse trains show additional amplitude modulation on the pulse train with period of 0.85 ns, corresponds to the frequency splitting near the external cavity mode at 1.18 GHz.

TABLE I. Parameters used in Eq. (1)–(8) for the THz QCL model.

| Symbol | Description | Value / Units |
|---|---|---|
| $I$ | Drive current | 1000 mA ($1.3I_{th}$) |
| $L$ | Laser cavity length | 2 mm |
| $n$ | Effective refractive index | 3.6 |
| $\Gamma_c$ | Confinement factor | 0.13 |
| $M$ | Number of stages | 90 |
| $V$ | Volume of active region | $3.6 \times 10^6 \ \mu m^3$ |
| $\omega_0$ | Central emission angular frequency | $2\pi \times 3$ THz |
| $\alpha$ | Linewidth enhancement factor | $-0.1$ |
| $\omega_{hom}$ | FWHM of gain spectrum | 400 GHz |
| $\varepsilon_0$ | Vacuum permittivity | $8.854 \times 10^{-12}$ F/m |
| $\varepsilon_b$ | Relative dielectric constant of QCL medium | 12.96 |
| $\tau_e$ | Carrier lifetime | 5 ps |
| $\tau_{ext}$ | Round-trip time, $\tau_{ext} = \frac{2L_{ext}n_{ext}}{c}$ | 0.67 ns |
| $L_{ext}$ | External cavity length | 0.1 m |
| $n_{ext}$ | Refractive index of external cavity | 1.00 |
| $\kappa$ | Feedback coupling coefficient | Varies |
| $\widetilde{\kappa}$ | Feedback coupling rate | Varies |
| $\varepsilon$ | Total coupling coefficient | Varies |
| $\varepsilon_L$ | Total external-cavity loss | 1 |
| $\varepsilon_S$ | Re-injection coupling factor | 1 |
| $t_L$ | Laser-facet transmissivity | 0.8246 |
| $R$ | Laser facet reflectivity | 0.32 |
| $R_{ext}$ | External target reflectivity | 0–100% |
| $C$ | Feedback parameter, $C = \widetilde{\kappa}\tau_{ext}\sqrt{1+\alpha^2}$ | Varies |
| $q$ | Elementary charge | $1.602 \times 10^{-19}$ C |
| $c$ | Speed of light | $2.99792458 \times 10^8$ m/s |

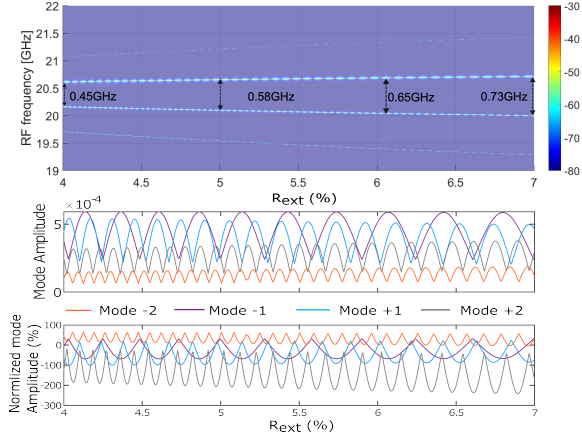

FIG. 3. RF beat note map (top) and mode amplitude variations (bottom) with the external reflectivity $R_{ext}$, which is varied from 4.0% to 7.0%.

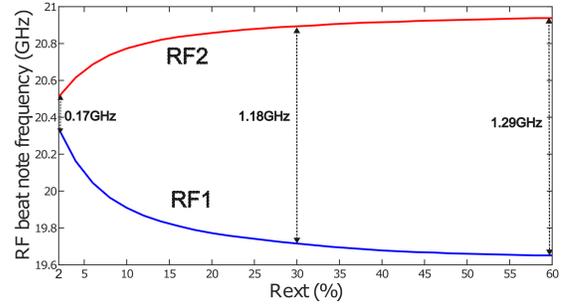

FIG. 4. The split two frequencies RF1 and RF2 from the RF beat note map as a function of the external reflectivity $R_{ext}$ varying between 2% to 60%.

To provide a complete dynamical study of the THz frequency comb, we also plot the RF beat note map under stronger feedback as $R_{ext}$ increases from 60% to 100% ($C > 13.86$, the value corresponding to $R_{ext} = 60\%$). Interestingly, although the frequency splitting dynamics continues in this region (Fig. 6, top), we also observe non-smooth oscillations in the mode amplitudes (Fig. 6, bottom). This is caused by spectral broadening of each comb line in the optical spectrum, which produces a more complex beating structure in the RF spectrum. Nevertheless, no fully complex or irregular RF beat note map appears in the strong feedback regime.

It is noteworthy that when the driving current of the THz QCL is increased from $1.3I_{th}$ (1000 mA) to $1.859I_{th}$ (1430 mA), the feedback dynamics transition from frequency combs to chaos output, as shown in Fig. 7. The time trace, optical spectrum and power spectrum at $R_{ext} = 0.8\%$ and 2% are compared. At $R_{ext} = 0.8\%$, the time trace exhibits a stable, constant envelope, whereas at $R_{ext} = 2\%$ it develops a random, irregular envelope indicative of chaos. Correspondingly, each comb line in the optical spectrum and the RF power spectrum broaden into continuous, noise-like distributions in the chaotic state. The chaotic state is further confirmed by the positive largest Lyapunov exponent (0.61227), calculated using Wolf's algorithm[42].

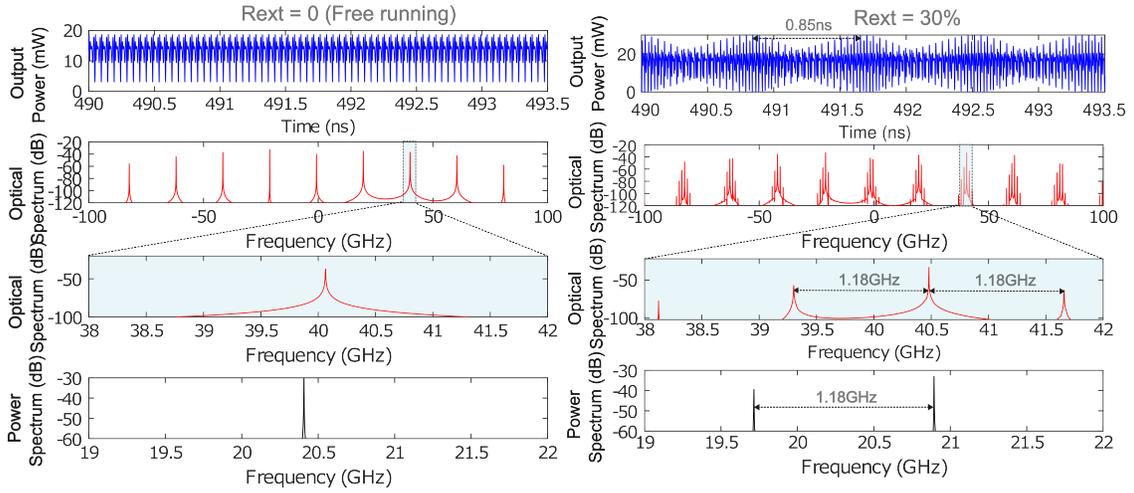

FIG. 5. Free-running output performance (left) compared with the output under optical-feedback dynamics with an external reflectivity of $R_{ext} = 30\%$ (right), including the time traces of the output power, optical spectrum, and the power spectrum.

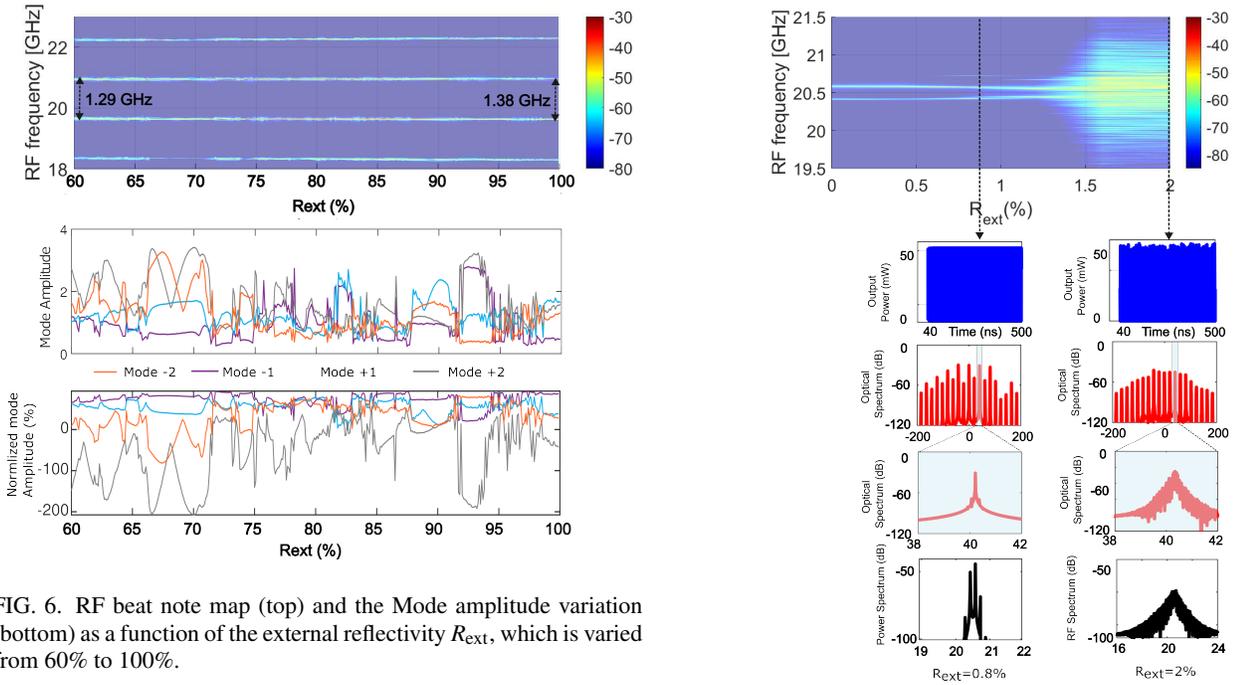

FIG. 6. RF beat note map (top) and the Mode amplitude variation (bottom) as a function of the external reflectivity $R_{ext}$, which is varied from 60% to 100%.

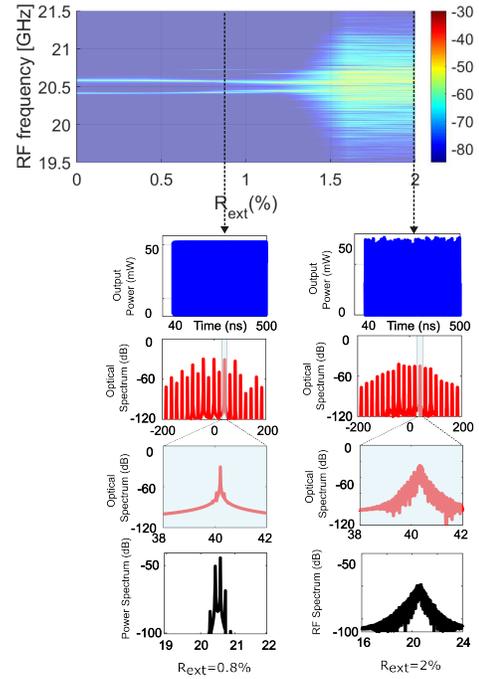

FIG. 7. RF beat note map as a function of the external reflectivity $R_{ext}$ varying between 0% to 2% at bias current 1430 mA. The laser output dynamics at $R_{ext} = 0.8\%$ and 2% are shown in the insets.

In Fig. 8, we investigate the influence of the linewidth enhancement factor $\alpha$ on the optical feedback dynamics of the THz QCL, with the driving current fixed at 1000 mA. Results are shown for $\alpha$ values ranging from −0.1 to 1.6, covering the range typically observed in THz QCLs, under a fixed reflectivity $R_{ext}$ at 1.45%. It can be observed that increasing $\alpha$ leads to linewidth broadening at each comb frequencies, which is also reflected in the resulting RF beat note spectrum. The largest Lyapunov exponent for each case, calculated using the Wolf's algorithm, is presented in Fig. 9. The Lyapunov exponent increases from values close to zero for $\alpha < 0.5$, to 0.382 at $\alpha = 0.5$, and further to 0.644 at $\alpha = 1.6$, indicating increasingly stronger chaotic behavior at higher $\alpha$.

## IV. CONCLUSION

Chaotic light is typically generated in diode lasers operating in the near-infrared or in MIR-QCLs under external per-turbations such as optical feedback. In this work, we demonstrate the optical feedback dynamics in a THz QCL exhibiting a free-running optical frequency comb (FC) output using Effective semiconductor Maxwell-Bloch equations. Notably, we obtain interesting bifurcation transitions in the weak feedback regime at low bias current (1000 mA), evolving from a stable FC to FC with P1, P2, P4, and P8 dynamics as the feedback level increases. For the excited P1 frequency, it initially appears at 0.15 GHz, approximately 10 times smaller than the resonant external cavity frequency, and increases with feedback from 2% to 60%, reaching 1.38 GHz, which is close to the 1.5 GHz resonant frequency of a 10 cm external cavity. Furthermore, routes to chaos are observed at higher bias currents ( 1430 mA) as feedback reflectivity increases. Calculations of the largest Lyapunov exponent show that the exponent increases alongside the bandwidth of the FC lines for each longitudinal mode with increasing linewidth enhancement factor. Simulations across various bias currents indicate that multi-mode emission at relatively high currents is necessary to generate chaos in a THz QCL with a low linewidth enhancement factor (−0.1). In contrast, chaotic regions are difficult to achieve under single-mode emission at low currents (< 800 mA), even with increased the external cavity lengths and higher feedback coefficients. This study of chaotic light generation in THz QCLs under optical feedback opens opportunities to explore typical chaos-based applications, such as LiDAR systems and secure communications, in the far-infrared wavelength range, which may offer improved performance compared to near- and mid-infrared bands, such as higher resolution for sensing and wider data bandwidth for communication systems.

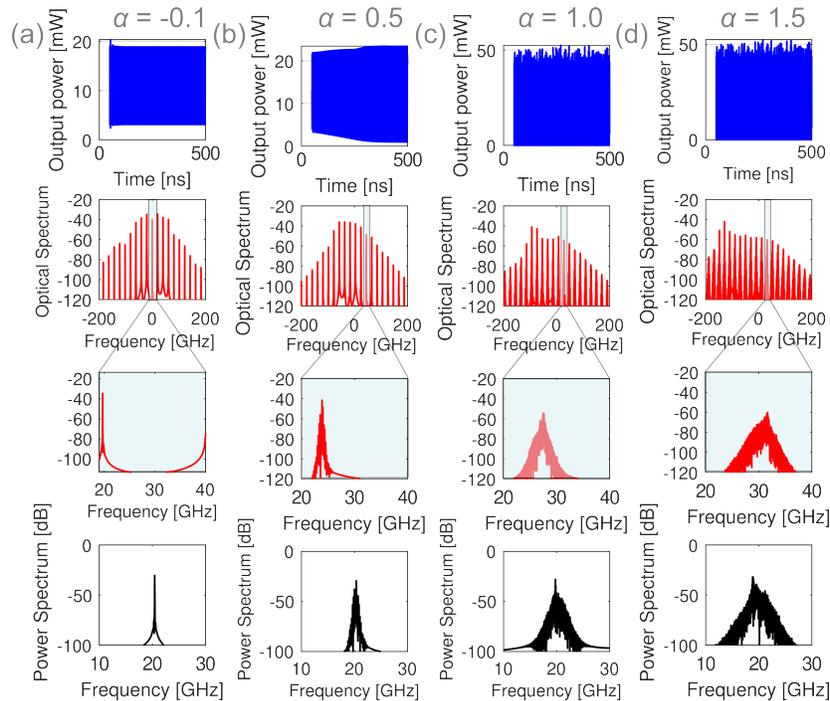

FIG. 8. Influence of the linewidth enhancement factor of the THz QCL α (increasing from −0.1, to 0.5, 1.0 and 1.5) on the optical feedback dynamics with $R_{ext}$ = 1.45%, including the time traces of the output power, optical spectrum and the power spectrum.

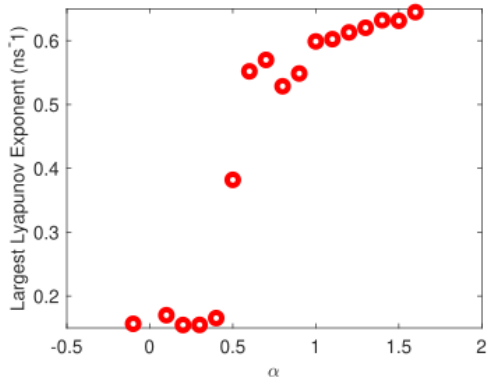

FIG. 9. The largest Lyapunov exponent as a function of the linewidth enhancement factor α for $R_{ext}$ = 1.45%.

## REFERENCES

[1]T. H. Maiman, R. H. Hoskins, I. J. D'Haenens, C. K. Asawa, and V. Evtuhov. Stimulated optical emission in fluorescent solids. ii. spectroscopy and stimulated emission in ruby. *Phys. Rev.*, 123:1151–1157, Aug 1961.

[2]M. Sciamanna and K. A. Shore. Physics and applications of laser diode chaos. *Nature Photonics*, 9(3):151–162, 2015.

[3]H Haken. Analogy between higher instabilities in fluids and lasers. *Physics Letters A*, 53(1):77–78, 1975.

[4]D. Lenstra, B. Verbeek, and A. Den Boef. Coherence collapse in single-mode semiconductor lasers due to optical feedback. *IEEE Journal of Quantum Electronics*, 21(6):674–679, 1985.

[5] J. Mørk, J. Mark, and B. Tromborg. Route to chaos and competition between relaxation oscillations for a semiconductor laser with optical feedback. *Phys. Rev. Lett.*, 65:1999–2002, Oct 1990.

[6] Yao Huang Kao, Nien Ming Wang, and Hong Ming Chen. Mode description of routes to chaos in external-cavity coupled semiconductor lasers. *IEEE Journal of Quantum Electronics*, 30(8):1732–1739, 1994.

[7] T. Heil, I. Fischer, W. Elsasser, and A. Gavrielides. Dynamics of semiconductor lasers subject to delayed optical feedback: the short cavity regime. *Physical review letters*, 87:243901, Dec 2001.

[8] T. Erneux, A. Gavrielides, and M. Sciamanna. Stable microwave oscillations due to external-cavity-mode beating in laser diodes subject to optical feedback. *Phys. Rev. A*, 66:033809, Sep 2002.

[9] L. Jumpertz, M. Carras, K. Schires, and F. Grillot. Regimes of external optical feedback in 5.6 um distributed feedback mid-infrared quantum cascade lasers. *Applied Physics Letters*, 105(13):131112, 10 2014.

[10] T. B. Simpson, J. M. Liu, A. Gavrielides, V. Kovanis, and P. M. Alsing. Period?doubling route to chaos in a semiconductor laser subject to optical injection. *Applied Physics Letters*, 64(26):3539–3541, 06 1994.

[11] S. Wieczorek, B. Krauskopf, T.B. Simpson, and D. Lenstra. The dynamical complexity of optically injected semiconductor lasers. *Physics Reports*, 416(1):1–128, 2005.

[12] Xiaoqiong Qi and Jia-Ming Liu. Dynamics scenarios of dual-beam optically injected semiconductor lasers. *IEEE Journal of Quantum Electronics*, 47(6):762–769, 2011.

[13] Mohammad AlMulla, Xiaoqiong Qi, and Jia-Ming Liu. Dynamics maps and scenario transitions for a semiconductor laser subject to dual-beam optical injection. *IEEE Journal of Selected Topics in Quantum Electronics*, 19(4):1501108–1501108, 2013.

[14] H.-F. Liu and W.F. Ngai. Nonlinear dynamics of a directly modulated 1.55 mu m ingaasp distributed feedback semiconductor laser. *IEEE Journal of Quantum Electronics*, 29(6):1668–1675, 1993.

[15] M. Sciamanna, A. Valle, P. Mégret, M. Blondel, and K. Panajotov. Nonlinear polarization dynamics in directly modulated vertical-cavity surface-emitting lasers. *Phys. Rev. E*, 68:016207, Jul 2003.

[16] Fan yi Lin and Jia-Ming Liu. Nonlinear dynamics of a semiconductor laser with delayed negative optoelectronic feedback. *IEEE Journal of Quantum Electronics*, 39(4):562–568, 2003.

[17] S. Tang and J.M. Liu. Chaotic pulsing and quasi-periodic route to chaos in a semiconductor laser with delayed opto-electronic feedback. *IEEE Journal of Quantum Electronics*, 37(3):329–336, 2001.

[18] M San Miguel, Q Feng, and Jerome V Moloney. Light-polarization dynamics in surface-emitting semiconductor lasers. *Physical Review A*, 52(2):1728, 1995.

[19] Martin Virte, Krassimir Panajotov, Hugo Thienpont, and Marc Sciamanna. Deterministic polarization chaos from a laser diode. *Nature Photonics*, 7(1):60–65, 2013.

[20] A. Uchida, T. Heil, Yun Liu, P. Davis, and T. Aida. High-frequency broadband signal generation using a semiconductor laser with a chaotic optical injection. *IEEE Journal of Quantum Electronics*, 39(11):1462–1467, 2003.

[21] Fan-Yi Lin and Jia-Ming Liu. Chaotic lidar. *IEEE Journal of Selected Topics in Quantum Electronics*, 10(5):991–997, 2004.

[22] Yuncai Wang, Bingjie Wang, and Anbang Wang. Chaotic correlation optical time domain reflectometer utilizing laser diode. *IEEE Photonics Technology Letters*, 20(19):1636–1638, 2008.

[23] Roman Lavrov, Maxime Jacquot, and Laurent Larger. Nonlocal nonlinear electro-optic phase dynamics demonstrating 10 gb/s chaos communications. *IEEE Journal of Quantum Electronics*, 46(10):1430–1435, 2010.

[24] Yanhua Hong. Flat broadband chaos in mutually coupled vertical-cavity surface-emitting lasers. *IEEE Journal of Selected Topics in Quantum Electronics*, 21(6):652–658, 2015.

[25] Kazusa Ugajin, Yuta Terashima, Kento Iwakawa, Atsushi Uchida, Takahisa Harayama, Kazuyuki Yoshimura, and Masanobu Inubushi. Real-time fast physical random number generator with a photonic integrated circuit. *Optics Express*, 25(6):6511 – 6523, 2017. Cited by: 73; All Open Access, Gold Open Access.

[26] Hao Zhang, Shuiying Xiang, Yahui Zhang, and Xingxing Guo. Complexity-enhanced polarization-resolved chaos in a ring network of mutually coupled vertical-cavity surface-emitting lasers with multiple delays. *Appl. Opt.*, 56(24):6728–6734, Aug 2017.

[27] Wenhui Chen, Xiaoxin Mao, Junli Wang, Rong Zhang, Longsheng Wang, Zhiwei Jia, Pu Li, Anbang Wang, and Yuncai Wang. Optical chaos generation and applications. *Advanced Photonics Research*, 6(9):2500055, 2025.

[28] Shuangquan Gu, Kun Li, Pei Zhou, and Nianqiang Li. Parallel optical chaos generation and ultrafast photonic decision-making based on a single quantum dot spin-vcsel. *Chaos, Solitons & Fractals*, 191:115874, 2025.

[29] L Jumpertz, F Michel, R Pawlus, W Elsässer, K Schires, M Carras, and F Grillot. Measurements of the linewidth enhancement factor of mid-infrared quantum cascade lasers by different optical feedback techniques. *AIP Advances*, 6(1):015212, 2016.

[30] Binbin Liu, Carlo Silvestri, Kang Zhou, Xuhong Ma, Shumin Wu, Ziping Li, Wenjian Wan, Zhenzhen Zhang, Ying Zhang, Junsong Peng, Heping Zeng, Cheng Wang, Massimo Brambilla, Lorenzo Luigi Columbo, and Hua Li. Terahertz semiconductor laser chaos. *Nature Communications*, 16(1):9985, 2025.

[31] F. P. Mezzapesa, L. L. Columbo, M. Brambilla, M. Dabbicco, S. Borri, M. S. Vitiello, H. E. Beere, D. A. Ritchie, and G. Scamarcio. Intrinsic stability of quantum cascade lasers against optical feedback. *Opt. Express*, 21(11):13748–13757, Jun 2013.

[32] L.L. Columbo and Massimo Brambilla. Multimode regimes in quantum cascade lasers with optical feedback. *Optics Express*, 22, 04 2014.

[33] Xiaoqiong Qi, Karl Bertling, Thomas Taimre, Gary Agnew, Yah Leng Lim, Tim Gillespie, Ashley Robinson, Michael Brünig, Aleksandar Demić, Paul Dean, Lian He Li, Edmund H. Linfield, A. Giles Davies, Dragan Indjin, and Aleksandar D. Rakić. Observation of optical feedback dynamics in single-mode terahertz quantum cascade lasers: Transient instabilities. *Phys. Rev. A*, 103:033504, Mar 2021.

[34] Xiaoqiong Qi, Karl Bertling, Thomas Taimre, Gary Agnew, Yah Leng Lim, Tim Gillespie, Aleksandar Demić, Paul Dean, Lian He Li, Edmund H. Linfield, A. Giles Davies, Dragan Indjin, and Aleksandar D. Rakić. Terahertz quantum cascade laser under optical feedback: effects of laser self-pulsations on self-mixing signals. *Opt. Express*, 29(24):39885–39894, Nov 2021.

[35] Xiaoqiong Qi, Karl Bertling, Thomas Taimre, Yah Leng Lim, Tim Gillespie, Paul Dean, Lian He Li, Edmund H. Linfield, A. Giles Davies, Dragan Indjin, and Aleksandar D. Rakić. Terahertz imaging with self-pulsations in quantum cascade lasers under optical feedback. *APL Photonics*, 6(9):091301, 09 2021.

[36] Ali Khalatpour, Andrew K. Paulsen, Chris Deimert, Zbig R. Wasilewski, and Qing Hu. High-power portable terahertz laser systems. *Nature Photonics*, 15(1):16–20, 2021.

[37] Xing-Guang Wang, Bin-Bin Zhao, Yu Deng, Vassilios Kovanis, and Cheng Wang. Nonlinear dynamics of a quantum cascade laser with tilted optical feedback. *Phys. Rev. A*, 103:023528, Feb 2021.

[38] Carlo Silvestri, Xiaoqiong Qi, Thomas Taimre, and Aleksandar D. Rakić. Frequency combs induced by optical feedback and harmonic order tunability in quantum cascade lasers. *APL Photonics*, 8(11):116102, 11 2023.

[39] Michael J. Wishon, Alexandre Locquet, C. Y. Chang, D. Choi, and D. S. Citrin. Crisis route to chaos in semiconductor lasers subjected to external optical feedback. *Phys. Rev. A*, 97:033849, Mar 2018.

[40] Carlo Silvestri, Lorenzo Luigi Columbo, Massimo Brambilla, and Mariangela Gioannini. Coherent multi-mode dynamics in a quantum cascade laser: amplitude- and frequency-modulated optical frequency combs. *Opt. Express*, 28(16):23846–23861, Aug 2020.

[41] Yah Leng Lim, Karl Bertling, Thomas Taimre, Tim Gillespie, Chris Glenn, Ashley Robinson, Dragan Indjin, Yingjun Han, Lianhe Li, Edmund H Linfield, et al. Coherent imaging using laser feedback interferometry with pulsed-mode terahertz quantum cascade lasers. *Optics express*, 27(7):10221–10233, 2019.

[42] Alan Wolf, Jack B. Swift, Harry L. Swinney, and John A. Vastano. Determining lyapunov exponents from a time series. *Physica D: Nonlinear Phenomena*, 16(3):285–317, 1985.